\documentclass[aps,prd,showpacs,preprintnumbers,twocolumn]{revtex4}
\usepackage{graphicx}
\usepackage{epsfig}
\usepackage{color}
\usepackage{amsmath}
\usepackage{epsf,amssymb,amsfonts,amsmath}

\def\be{\begin{equation}}
\def\ee{\end{equation}}
\def\bea{\begin{eqnarray}}
\def\eea{\end{eqnarray}}

\newcommand{\beq}{\begin{equation}}
\newcommand{\eeq}{\end{equation}}
\newcommand{\beqa}{\begin{eqnarray}}
\newcommand{\eeqa}{\end{eqnarray}}
\newcommand{\beqar}{\begin{eqnarray*}}
\newcommand{\eeqar}{\end{eqnarray*}}

\renewcommand{\eqref}[1]{(\ref{#1})}

\catcode`\@=12

\begin{document}

\title{Realization of chiral symmetry breaking and restoration in holographic QCD}
\author{Kaddour Chelabi $^{a,b,c}$}
\author{Zhen Fang$^{a,b}$}
\author{Mei Huang$^{d,e}$}
\author{Danning Li$^{a}$}
\author{Yue-Liang Wu$^{a,b}$}
\affiliation{$^{a}$ State Key Laboratory of Theoretical Physics,
Institute of Theoretical Physics, Chinese Academy of Sciences,
Beijing 100190, P. R. China }
\affiliation{$^{b}$ University of Chinese Academy of Sciences (UCAS), P.R. China}
\affiliation{$^{c}$ Laboratory of Particle Physics and Statistical Physics, Ecole Normale Superieure-Kouba. B.P. 92,16050, Vieux-Kouba, Algiers, Algeria}
\affiliation{$^{d}$Institute of High
Energy Physics, Chinese Academy of Sciences, Beijing 100049, P.R. China}
\affiliation{$^{e}$Theoretical Physics Center for Science
Facilities, Chinese Academy of Sciences, Beijing 100049, P.R. China}

\begin{abstract}
With proper profiles of the scalar potential and the dilaton field,  for the first time, the spontaneous chiral symmetry breaking
in the vacuum and its restoration at finite temperature are correctly realized in the holographic QCD framework.
In the chiral limit, a nonzero chiral condensate develops in the vacuum and decreases with temperature, and the phase transition
is of 2nd order for two-flavor case and of 1st order for three-flavor case. In the case of explicit chiral symmetry breaking,
in two-flavor case, the 2nd order phase transition turns to crossover with any nonzero current quark mass, and in three-flavor case,
the 1st order phase transition turns to crossover at a finite current quark mass. The correct description of chiral symmetry breaking
and restoration makes the holographic QCD models more powerful in dealing with non-perturbative QCD phenomena.
This framework can be regarded as a general set up in application of AdS/CFT to describe conventional Ginzburg-Landau-Wilson
type phase transitions, e.g. in condensed matter and cosmology systems.
\end{abstract}
\pacs{13.40.-f, 25.75.-q, 11.10.Wx }
\maketitle

{\it Introduction:} The spontaneous chiral symmetry breaking and the color charge confinement are two most intriguing
non-perturbative aspects of
Quantum Chromodynamics (QCD). It is widely believed that chiral symmetry can be restored and color degrees of freedom can
be freed at high temperature and/or density. The interplay between chiral and deconfinement phase transitions are of continuous interests for
studying the QCD phase diagram, which is always the most important topic of high energy nuclear physics. The deconfined quark-gluon matter
might be formed during the evolution of the early universe at high temperature, and could exist in the core of compact stars at high baryon
density, respectively. In laboratory, ultrarelativistic heavy ion collisions (HIC) at the Relativistic Heavy Ion collider (RHIC) and the Large
Hadron Collider (LHC) provide a unique controllable experimental tool to investigate QCD phase transitions and properties of quark matter
at high temperature and small baryon density, while future facilities at FAIR and NICA will focus on exploring QCD phase structures at finite
baryon density.

The chiral symmetry breaking and restoration is well defined in the chiral limit when the current quark mass is zero, and characterized by the
order parameter $\langle{\bar q}q \rangle$, i.e., the chiral condensate \cite{Nambu:1961tp,Nambu:1961fr}.
The confinement deconfinement phase transition is related to the center
symmetry, which is only well defined in pure gauge sector when the current quark mass goes to infinity $m\rightarrow \infty$, and the order
parameter is characterized by the Polyakov loop expectation value $\langle L \rangle $ \cite{Polyakov:1978vu}. In principle,  the chiral restoration and deconfinement phase transition can be separated. At zero density, lattice QCD results show that in the chiral limit the chiral and deconfinement phase transitions occur at the same critical temperature \cite{Kogut:1982rt}. However, in the case of physical quark mass, lattice simulations \cite{nature-PTD} shows that the critical temperatures for chiral restoration of light quarks $T_c^{\chi,(u,d)}=151 \text{MeV}$ and deconfinement phase transitions $T_c^d=176 \text{MeV}$ are different. It is also known that
\cite{qcd-phase-diagram,deForcrand:2006pv,Kanaya:2010qd,Pisarski:1983ms}, in the chiral limit, the chiral phase transition is of second order phase transition for two-flavor case, and is of first order for three-flavor case due to the three-flavor mixing term or t'Hooft determinant term. When considering the finite quark mass, the chiral phase transition turns to be a crossover.

QCD vacuum properties, QCD phase transitions and hot/dense matter around critical temperature/density are mainly dominated by
non-perturbative dynamics, hence perturbative methods become invalid in this region. Non-perturbative methods such as lattice QCD,
Dyson-Schwinger equations (DSEs), and functional renormalization group equations (FRGs) have been developed for several decades.
Recently, the conjecture of the gravity/gauge duality \cite{Maldacena:1997re,Gubser:1998bc,Witten:1998qj} provides a new hopeful
tool to tackle the strong coupling problem of the strong interaction. By breaking the conformal symmetry in different ways, many efforts
have been made both in top-down and bottom-up approaches towards more realistic holographical description of QCD in non-perturbative
region, such as hadron physics, QCD phase transitions, thermodynamical and transport properties of hot/dense QCD matter  (see \cite{Aharony:1999ti,Erdmenger:2007cm,CasalderreySolana:2011us,deTeramond:2012rt,Adams:2012th}
 for reviews).

For QCD phase transitions, most bottom-up holographic studies focus on confinement/deconfinement phase transition related to pure gauge sector.
However, the chiral phase transition has been much less studied in the framework of bottom-up holography due to the complexity when adding
flavor dynamics. In top-down approach, normally $D_p-D_q$ system is used in type II superstring theory, with the $D_p$ background
brane describing the effects of pure QCD non-Abelian gauge theories and the $D_q$ probe brane describing the flavor dynamics, e.g.
Sakai-Sugimoto model(D4-D8)\cite{SS-1,SS-2} and D3-D7\cite{D3-D7}  systems.
 In bottom-up approach, normally a flavor action is added on the metric background
as in the hard-wall  \cite{Erlich:2005qh} and soft-wall
\cite{Karch:2006pv,TB:05,Colangelo:2011sr,Gherghetta-Kapusta-Kelley,Gherghetta-Kapusta-Kelley-2,YLWu,YLWu-1,Cui:2013xva,Li:2012ay,Li:2013oda}
holographic QCD models. The chiral phase transition of QCD has never been correctly realized in the framework of AdS/CFT from both top-down
and bottom-up approaches. In this work, we show how to correctly realize chiral symmetry breaking in the vacuum and restoration at finite temperature
in the soft-wall holographic QCD models.

In bottom-up approaches, hard-wall model \cite{Erlich:2005qh} and soft-wall model \cite{Karch:2006pv} are successful in describing hadron physics. Their extended models
\cite{Gherghetta-Kapusta-Kelley,Gherghetta-Kapusta-Kelley-2,YLWu,YLWu-1,Cui:2013xva,Li:2012ay,Li:2013oda}
can describe the hadron spectra and related quantities in very good accuracy. In these models, the chiral condensate is introduced to realize the spontaneous chiral symmetry breaking at zero temperature. However, there is no correct quark mass dependence for the chiral condensate, and cannot give the correct chiral symmetry breaking mechanism at zero temperature. Furthermore, the chiral phase transition at finite temperature cannot be correctly realized in the soft-wall holographic models. We find that both profiles of the scalar potential and the dilaton field are essential to generate correct quark mass dependence behavior of the chiral condensate.

{\it The framework:} In this work, we only focus on chiral symmetry breaking and restoration, therefore, we only take the scalar part of the
$SU(N_f)_L\times SU(N_f)_R$ 5D action, which takes the following form
\begin{eqnarray}\label{action}
 S=-\int d^5x
 \sqrt{-g}e^{-\Phi}Tr(D_m X^+ D^m X+V_X(|X|)).
\end{eqnarray}
Here $\Phi$ is the dilaton field, and the 5D mass of the complex scalar field $X$ can be determined as $M_5^2=-3$ from the AdS/CFT dictionary $M_5^2=(\Delta-p)(\Delta+p-4)$\cite{Witten:1998qj} (here we have set the AdS radius $L=1$)  by taking $\Delta=3, p=0$.
In this work, we do not consider the back-reaction of the dilaton field and scalar field to the background geometry, and $g$ is the AdS$_5$ metric
background. In the string frame one has
\begin{eqnarray}\label{metric-ansatz}
d s^2=e^{2A_s(z)}(-f(z) d t^2+\frac{1}{f(z)}d z^2+dx_i dx^i),
\end{eqnarray}
with the background metric Eq.(\ref{metric-ansatz}) as the AdS-Schwarzchild black hole solution, in which
\begin{eqnarray}
A_s(z)&=&-\log(z),\label{adsas}\\
f(z)&=&1-\frac{z^4}{z_h^4}.\label{adsf}
\end{eqnarray}
Here $z_h$ is the horizon of the black hole defined at $f(z_h)=0$ and related to the temperature $T$ of the system by the Hawking formula
\begin{eqnarray}
T=|\frac{f^{'}(z_h)}{4\pi}|=\frac{1}{\pi z_h}.
\end{eqnarray}

If the scalar field $X$ gets a non-vanishing vacuum expectation value $X_0$, then $SU(N_f)_L\times SU(N_f)_R$ is spontaneously broken.
In this work, we would work in the case that $m_u=m_d$ for $N_f=2$ and $m_u=m_d=m_s$ for $N_f=3$, so we would expect that the symmetry would be broken to $SU(N_f)$ and $X_0=\frac{\chi(z)}{\sqrt{2N_f}}I_{N_f}$. Here $I_{N_f}$ is the $N_f\times N_f$ identity matrix and $\chi(z)$ is assumed to depend only on the
 fifth coordinate $z$. Inserting the expectation value of $X$, we get the effective description in terms of $\chi$ of the following form
\begin{eqnarray}\label{eff-action}
S_{\chi}=-\int d^5x
 \sqrt{-g}e^{-\Phi}(\frac{1}{2}g^{zz}\chi^{'2}+V(\chi)),
\end{eqnarray}
where we simply denote the scalar potential in terms of $\chi$ as
\begin{equation}\label{profile-chi}
V(\chi)\equiv Tr({V_X(|X|)})=-\frac{3}{2}\chi^2+v_3 \chi^3+v_4 \chi^4.
\end{equation}
 The leading term of $V(\chi)$ comes from the mass term and it is fixed to be $-\frac{3}{2}\chi^2$. The quartic term $v_4 \chi^4$ keeps $\chi\leftrightarrow-\chi$ symmetry, and the cubic term $v_3 \chi^3$ comes from the three-flavor mixing term and vanishes for the two-flavor case.
The equation of motion for $\chi$ with respect to the action Eq.(\ref{eff-action}) can be derived as
\begin{eqnarray}\label{eom-chi-1}
\chi^{''}+(3A_s^{'}-\Phi^{'}+\frac{f^{'}}{f})\chi^{'}- \frac{e^{2A_s}}{f}\partial_\chi V(\chi)=0.
\end{eqnarray}
According to the AdS/CFT prescription, the UV asymptotic behavior of $\chi(z)$, which is assumed to be dual to the $\bar{q}q$ operator in QCD,
and the leading UV expansion of this equation is still of the form $m_q \zeta z+\frac{\sigma}{\zeta} z^3$ with $m_q$ the current quark mass, $\sigma$ the chiral condensate and $\zeta=\frac{\sqrt{3}}{2\pi}$ \cite{Cherman:2008eh}.

In order to solve the chiral condensate from Eq. (\ref{eom-chi-1}), one has to know the profile of the dilaton field $\Phi(z)$. The positive
dilaton background as in the original soft-wall model cannot generate correct chiral symmetry breaking in the chiral limit.
We find that a negative dilaton field as introduced in \cite{Zuo:2009dz,Gutsche:2011vb} can create the chiral symmetry breaking in the
chiral limit and can realize chiral phase transition, for details see \cite{newlongpaper}. However, as pointed out in \cite{Karch:2006pv,KKSS-2},
the negative dilaton background predicts an un-physical massless scalar meson state which is unacceptable. The results from spectra analysis
and thermodynamical analysis seem in contradiction. To solve this issue, one has to note that the
dominating energy scale of chiral symmetry breaking is around $1{\rm GeV}$ and that of confinement is around $200\sim300{\rm MeV}$\cite{shuryak-twoscales}.
In the Regge behavior analysis, the confinement is the dominating effect while in chiral symmetry breaking mechanism new scale should be introduced. 
Taking into account the latter, we expect that the negative dilaton dominates at small $z$ and at large $z$ positive dilaton would dominate.
In between the two, we take the following simple interpolation
\begin{eqnarray}\label{int-dilaton}
\Phi(z)=-\mu_1z^2+(\mu_1+\mu_0)z^2\tanh(\mu_2z^2),
\end{eqnarray}
where $\mu_0=(0.43 {\rm GeV})^2$ taken from \cite{Li:2012ay,Li:2013oda} to produce
the Regge spectra. At ultraviolet (UV) when $z\rightarrow 0$, $\Phi(z)\rightarrow -\mu_1z^2$, and at infrared (IR) limit
when $z\rightarrow \infty$ the above interpolation goes to $\Phi(z)\rightarrow\mu_0z^2$, which is responsible for the linear confinement. In fact, QCD at IR region has actually no exact conformal symmetry. Therefore, we need to consider IR improved Soft-Wall AdS/QCD. In general, such an IR modified AdS/QCD model can include IR improved metric, quartic term, dilaton and conformal mass. In this paper, we shall focus on the simple IR improved dilaton that changes sign in an appropriate region as shown in Eq.(\ref{int-dilaton}). The other IR improved Soft-Wall AdS/QCD models will be investigated elsewhere.

\begin{figure}
 \centerline{\includegraphics[width=7.5cm,height=6cm]{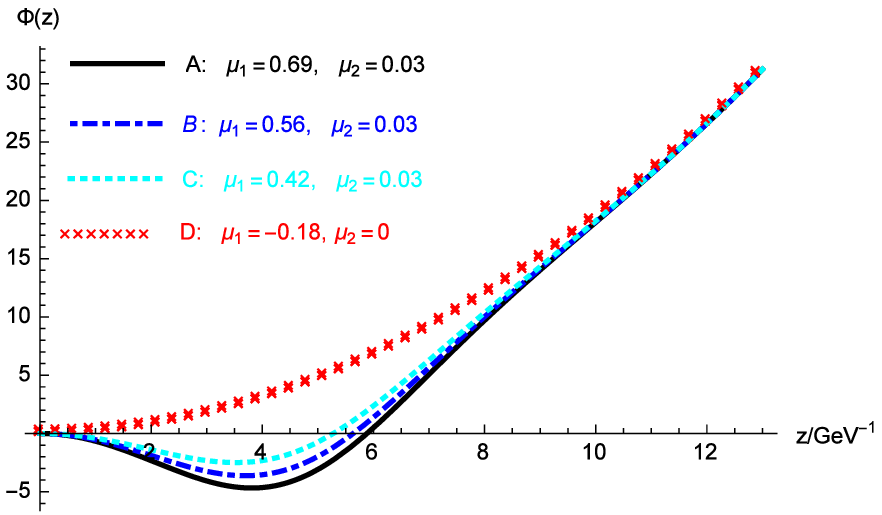}}
 \centerline{(a)}
\vfill
\centerline{\includegraphics[width=7.5cm,height=6cm]{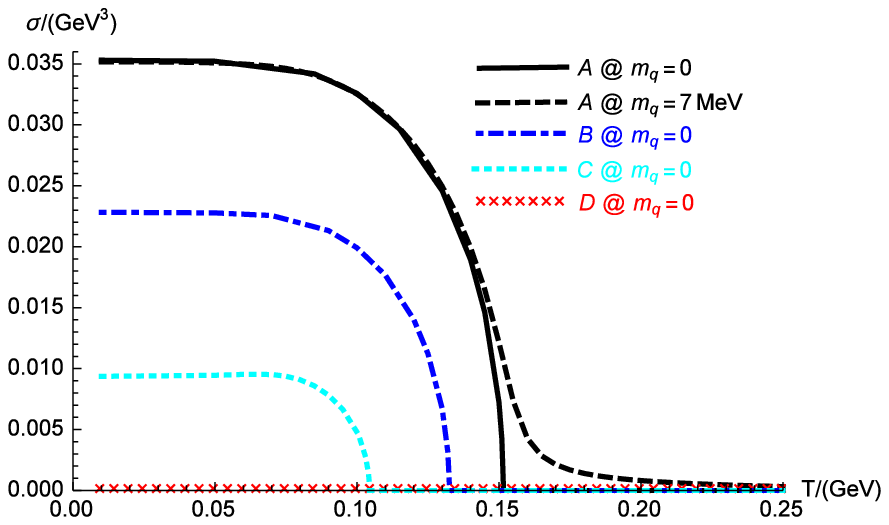}}
 \centerline{(b)}
\caption{(a) The profile of the dilaton field $\Phi(z)$ as a function of $z$ with different sets of parameters in unit of
${\rm GeV}^2$. (b) The corresponding solution of the chiral condensate $\sigma$
in two-flavor case as a function of the temperature $T$. }
 \label{dilatonprofile-chiralcondensate}
\end{figure}

We firstly consider the two-flavor $N_f=2$ case, where $v_3=0$ in Eq.(\ref{profile-chi}), we choose $v_4=8$ in order
to produce proper scalar mass as in \cite{YLWu}.
We show the dilaton field profile $\Phi(z)$ as a function of $z$ in Fig.\ref{dilatonprofile-chiralcondensate}(a),
and the corresponding solution of the chiral condensate $\sigma$ from Eq.(\ref{eom-chi-1}) as a function of the temperature in
Fig.\ref{dilatonprofile-chiralcondensate}(b).
It is noticed that if we choose the positive dilaton field profile (the red cross line) as in the original soft-wall model \cite{Karch:2006pv}, in the chiral limit, the chiral condensate $\sigma$ vanishes for all temperatures. This means there is no chiral symmetry breaking in the vacuum
with a positive dilaton field profile. When we choose the dilaton profile which is negative at UV and positive at IR as shown in black solid, blue dashed, and cyan dotdashed lines in Fig.\ref{dilatonprofile-chiralcondensate}(a),  a nonzero chiral condensate develops in the vacuum
as shown in Fig.\ref{dilatonprofile-chiralcondensate}(b). This means that the chiral symmetry is spontaneously broken in the vacuum.
When the temperature increases, the chiral condensate decreases and drops to zero at the critical temperature $T_c$, which shows that the chiral phase transition is of second order in the chiral limit. With proper choice of $\mu_1=(0.83 {\rm GeV})^2\simeq0.69 {\rm GeV}^2$  and $\mu_2=(0.176 {\rm GeV})^2\simeq 0.03 {\rm GeV}^2$, one can have the critical temperature $T_c=151 {\rm MeV}$. When the current quark mass is nonzero, e.g. $m_0=7 {\rm MeV}$, the chiral symmetry is explicitly broken,
we find that the phase transition turns to be a crossover as shown by the black dashed line in Fig.\ref{dilatonprofile-chiralcondensate}(b). With the same parameters choice of
$\mu_1= 0.69 {\rm GeV}^2$ and $\mu_2= 0.03 {\rm GeV}^2$, the pseudo-critical temperature is
around $T_c=150 {\rm MeV}$, which is in agreement with lattice results in \cite{nature-PTD}.

\begin{figure}
 \centerline{\includegraphics[width=7.5cm,height=6cm]{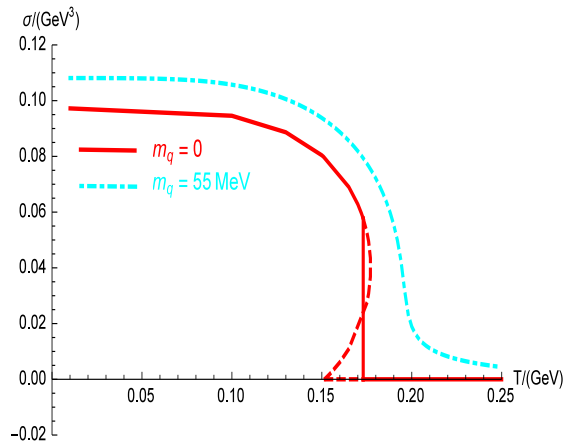}}
\caption{The chiral condensate $\sigma$ in three-flavor case as a function of temperature $T$ with the set of parameters:
$v_3=-3, v_4=8$, $\mu_1= 0.69 {\rm GeV}^2$ and $\mu_2= 0.03 {\rm GeV}^2$ with $m_q=0$ and $m_q=55 {\rm MeV}$,
respectively.}
 \label{1st-crossover}
\end{figure}

In the case of three-flavor ($N_f=3$), for simplicity, we take $m_q=m_u=m_d=m_s$, we have $v_3\neq 0$ due to the three-flavor
mixing term or the t'Hooft determinant term. We choose
$v_3=-3$ and keep the set of parameters $v_4=8$, $\mu_1= 0.69 {\rm GeV}^2$ and $\mu_2= 0.03 {\rm GeV}^2$ as in the
two-flavor case. In Fig.\ref{1st-crossover}, we show the chiral condensate as a function of temperature $T$ with $m_q=0$ and
$m_q=55 {\rm MeV}$, respectively. It is found that in the chiral limit, the temperature dependent chiral condensate shows a
typical 1st-order phase transition behavior (in red solid line and red dashed line), i.e., in certain temperature region,
there exist three solutions for the chiral condensate. In order to determine the stable solution, one has to compare the free energy
of the system, which is extracted from the on-shell action and takes the following form:
\begin{eqnarray}\label{freeenergy}
\mathcal{F}\equiv \frac{F}{V_3}
 &=&\int dz \sqrt{-g}e^{-\Phi}(-\frac{1}{2}v_3\chi^3-v_4 \chi^4)\nonumber\\
 & & -\frac{1}{2}(\chi e^{3A_s-\Phi}f\chi^{'})|_{\epsilon}.
\end{eqnarray}
When $m_q\neq0$, the last equation is divergent near $\epsilon=0$, and one has to add counter term to cancel the divergence, while when $m_q=0$, the UV divergence disappears and the last equation can be directly calculated after inserting the solutions of Eq.(\ref{eom-chi-1}). The minimum free energy
determines the stable solution of the chiral condensate at fixed temperature, and the stable solution of the chiral condensate
as the function of temperature is shown by red solid line in Fig.\ref{1st-crossover}. It is shown that at the critical temperature
$T_c=173 {\rm MeV}$, the chiral condensate jumps to zero and the phase transition is of 1st-order. In the case of explicit
chiral symmetry breaking with a finite current quark mass $m_q=55 {\rm MeV}$, the phase transition turns to be a cross-over.
This scenario is in agreement with the "Colombia Plot" in \cite{qcd-phase-diagram}.
We will consider the real three-flavor case with $m_u=m_d<m_s$ in the near future.

{\it Summary:} In this work, we investigate the chiral symmetry breaking and chiral phase transition in the framework
of soft-wall holographic QCD model.

In the original soft-wall model with positive dilaton field, the chiral condensate is introduced to characterize the chiral symmetry breaking.
However, it is worthy of mentioning that, in the chiral limit, the chiral condensate is zero at zero temperature, which means there is no
spontaneous chiral symmetry breaking in the vacuum in this model. In order to realize the chiral symmetry breaking in the chiral limit, we
find that one has to choose proper profiles for both the scalar potential and the dilaton field. Firstly, it is necessary to add the non-linear
terms in the scalar potential, and we assume the simplest non-linear potential by adding the cubic term and the quartic term, where the
cubic term only appears in three-flavor case due to the flavor mixing. Secondly, it is found that the positive dilaton background in the original
soft-wall model does not give correct behavior of chiral condensate even when taking into account the non-linear potential for the scalar field,
but a negative dilaton background can give prediction on chiral phase transition in good agreement with the Columbia sketch qualitatively.
Since chiral symmetry breaking and confinement happen in different scales, it is reasonable to assume the negative dilaton and positive dilaton
dominating in different regions. By interpolating both the negative dilaton at UV (determined by the chiral symmetry breaking) and the positive
quadratic dilaton at IR (determined by the linear confinement) in a simple way, one can avoid the massless scalar meson state and get a result qualitatively
similar to the negative dilaton background result.

With nonlinear scalar potential and the dilaton field which is negative at UV and  positive at IR, for the first time,
we can realize the spontaneous chiral symmetry breaking in the vacuum and its restoration at finite temperature
in the holographic QCD framework.  In the chiral limit, a nonzero chiral condensate develops in the vacuum and
decreases with the temperature, and the phase transition is of second order for two-flavor case and of 1st-order for
three-flavor case. In the case of finite current quark mass, the phase transition turns to crossover in both two-flavor and
three-flavor cases.

The profile of the scalar potential determines the possible solution structure of the chiral condensate, which is the sufficient condition
for the chiral condensate and determines the order of phase transition. The profile of the dilaton field is determined by gluodynamics,
which is the necessary condition for the chiral condensate. Our results indicate that only with proper gluodynamics running from UV to IR,
the chiral symmetry can be spontaneously broken in the vacuum, and chiral phase transition can happen at finite temperature.
The correct description of chiral symmetry breaking and restoration makes the holographic QCD models more powerful in dealing
with non-perturbative QCD phenomena, and the framework in this work can be regarded as a general set up in application of
AdS/CFT to describe conventional Ginzburg-Landau-Wilson type phase transitions, e.g. in condensed matter and cosmology systems.

In this work, we do not constrain the dilaton profile from the meson spectral simultaneously, and the back-reaction from the dilaton field
to the metric background is not considered,  which will be left for future work. Furthermore, after introducing chemical potential through
the gauge field part, we expect to get the correct $T-\mu$ phase diagram in the future.

\vskip 0.5cm
{\bf Acknowledgement}
\vskip 0.2cm
The authors thank Song He, Yi Yang for valuable discussions. K.C is supported by CAS-TWAS president fellowship.
M.H. is supported by the NSFC under Grant Nos. 11175251 and 11275213, DFG and NSFC (CRC 110),
CAS key project KJCX2-EW-N01, and Youth Innovation Promotion Association of CAS. This paper is funded in part by China Postdoctoral Science Foundation.

\end{document}